\documentclass[twocolumn,traditabstract]{aa}
\usepackage{graphicx}
\usepackage{amsbsy}
\usepackage{amsmath}
\usepackage{natbib}
\usepackage{color}
\usepackage{txfonts}
\usepackage{url}
\usepackage[usenames,dvipsnames,svgnames,table]{xcolor}

\begin{document}

\title{Ice condensation as a planet formation mechanism}
\titlerunning{Ice condensation as a planet formation mechanism}

\author{Katrin Ros \and Anders
Johansen}
\authorrunning{Ros \& Johansen} 

\institute{Lund Observatory, Department of Astronomy and Theoretical Physics,
Lund University, Box 43, 221 00 Lund, Sweden\\ \email{katrin.ros@astro.lu.se}}

\abstract{We show that condensation is an efficient particle growth mechanism, leading to growth beyond decimeter-sized pebbles close to an ice line in protoplanetary discs. As coagulation of dust particles is frustrated by bouncing and fragmentation, condensation could be a complementary, or even dominant, growth mode in the early stages of planet formation. Ice particles diffuse across the ice line and sublimate, and vapour diffusing back across the ice line recondenses onto already existing particles, causing them to grow. We develop a numerical model of the dynamical behaviour of ice particles close to the water ice line, approximately 3 AU from the host star. Particles move with the turbulent gas, modelled as a random walk. They also sediment towards the midplane and drift radially towards the central star. Condensation and sublimation are calculated using a Monte Carlo approach. Our results indicate that, with a turbulent $\alpha$-value of 0.01, growth from millimeter to at least decimeter-sized pebbles is possible on a time scale of 1000 years. We find that particle growth is dominated by ice and vapour transport across the radial ice line, with growth due to transport across the atmospheric ice line being negligible. Ice particles mix outwards by turbulent diffusion, leading to net growth across the entire cold region. The resulting particles are large enough to be sensitive to concentration by streaming instabilities, and in pressure bumps and vortices, which can cause further growth into planetesimals. In our model, particles are considered to be homogeneous ice particles. Taking into account the more realistic composition of ice condensed onto rocky ice nuclei might affect the growth time scales, by release of refractory ice nuclei after sublimation. We also ignore sticking and fragmentation in particle collisions. These effects will be the subject of future investigations.}


\maketitle

\section{Introduction}

Planets form in protoplanetary discs of gas and dust, surrounding young stars. In the classical planet formation scenario dust grains collide, stick together and form larger and larger bodies \citep{safronov1969}. Particles of up to millimeter-sizes stick together due to contact forces and kilometer-sized and larger bodies are held together by gravity. How particles grow from millimeter to kilometer-sized planetesimals is difficult to explain by collisions, as these particles tend to fragment or bounce off each other as they collide, instead of sticking \citep{blumwurm2008}. Further, solids on Keplerian orbits experience a headwind from the slow-orbiting pressure-supported gas, and therefore lose angular momentum and drift in towards the central star \citep{weidenschilling1977}. This radial drift velocity peaks for meter-sized particles, which drift into the star from 1 AU on a time-scale of only 100 orbits \citep{braueretal2008}.

Once particles grow to centimeter-sized to decimeter-sized pebbles, further growth into planetesimals is possible via particle concentration and gravitational collapse. The streaming instability causes particles to clump together due to the velocity difference between the particles and the gas in the disc, and planetesimals form through subsequent gravitational collapse \citep{youdingoodman2005, johansenetal2007}. Particles also concentrate in long-lived vortices \citep{bargesommeria1995, klahrbodenheimer2003} and pressure bumps \citep{johansenyoudinklahr2009} excited in the turbulent flow. However, an efficient mechanism for growth from millimeter-sized dust grains to pebbles is needed to explain the formation of particles that are large enough to take part in such concentration events.

Sticking, or coagulation, as a growth mechanism has been extensively studied, both experimentally \citep{blumwurm2008, zsometal2010} and numerically \citep{braueretal2008, windmarketal2012}. However, an often overlooked concept in planet formation models is that of growth via condensation near ice lines. \citet{stevensonlunine1988} investigated material enhancement at the ice line due to diffusive redistribution and subsequent condensation of water vapour from the inner part of the disc. With the assumptions of efficient condensation by homogeneous nucleation, and by ignoring inwards transport of material, the authors found a significant solid density enhancement. In contrast, \citet{cuzzizahnle2004} considered inwards material drift across the ice line, which would enhance the vapour density in the inner disc, followed by a phase of accumulation of solids by condensation onto immobile planetesimals formed outside of the ice line. 

In this work the dynamical behavior and growth of small ($\sim$1 mm) seed particles via ice condensation is studied in computer simulations. Ice particles moving with the turbulent gas encounter the water ice line where ice sublimates and becomes water vapour. We consider both the {\it radial ice line}, separating the inner, hotter part of the disc from the colder region further away from the star, and the {\it atmospheric ice line}, which separates the hot atmosphere of the disc from the cold midplane layer \citep{dullemonddominik2004, meijerinketal2009}. Due to the turbulent motion of the gas in the protoplanetary disc water vapour diffuses across the ice line into the condensation region where it condenses onto existing ice particles. Many ice particles will move across the ice line and sublimate. However, since small particles are coupled to the gas and thereby effectively move in a random walk due to turbulence, a significant fraction will be lucky enough to stay within the condensation region of the disc, growing to larger sizes as diffusing water vapour condenses onto them. The total mass in the ice component is maintained during this process, as the loss of ice particles to sublimation is completely compensated for by the growth of the remaining particles. 

In a moderately turbulent disc we find particle growth to beyond decimeter-sizes on a time-scale of a few 1000 orbits. The most significant growth takes place locally, close to the radial ice line. However, efficient radial mixing supplies the midplane with large particles even several scale heights outside of the ice line. The atmospheric ice line, located at $z\gtrsim3\,H$ where disc material is heated directly by the central star \citep{chianggoldreich1997}, contributes less to the particle growth, generally on the order of $\rm \mu m$.

The radial water ice line, or condensation front, is found at $r\approx3\,\rm AU$ around a solar type star \citep{lecaretal2006}. However, growth by condensation is not only applicable to water ice, but also to any other volatile found in the protoplanetary disc. Other important condensation fronts relevant for planet formation include those of ammonia, methane, carbon monoxide and molecular nitrogen at much larger radii than the water ice line, and of silicates at $r\ll 1\rm \, AU$ \citep{lodders2003, qietal2011}. These condensation fronts will be the topic of a future study.

The paper is organised as follows. In Section 2 the numerical model used for  particle dynamics and growth by condensation is described, including the units used in this paper. The Monte Carlo scheme for condensation is further explained in Section 3. In Section 4 we describe the test problems used to validate the code. We present our results in Section 5, and investigate the influence of the position of the atmospheric ice line, the turbulence strength and the effects of radial mixing. In a moderately turbulent disc ($\alpha=10^{-2}$), pebbles of centimeters to decimeters form within a few $1000$ years. We discuss general assumptions and simplifications made in Section 6. Finally, in Section 7 we conclude that particle growth by condensation is an important mechanism for forming pebbles in protoplanetary discs, which should be taken into account in future studies of planetesimal formation.

\section{Numerical model}
\label{sec:model}

\subsection{Simulation box, units and boundary conditions}

We simulate a two-dimensional region around the water ice line. The simulation domain is set in the radial $r$ and vertical $z$ direction, whereas the azimuthal direction is ignored, for simplicity and under the assumption of axisymmetry. The fundamental units are the gas scale height $H$ as a length unit, sound speed $c_{\rm s}$ as a velocity unit and inverse Keplerian orbits $\varOmega^{-1}$ as a time unit. The relation between these quantities is $H=c_{\rm s}/\varOmega$. Setting $H=c_{\rm s}=\varOmega=1$ gives a system which is scalable to any region of the protoplanetary disc. Hence our results apply equally well to other condensation fronts, such as those of CO or silicates, although the transition and scaling between drag regimes and the fractional ice abundance depend on the choice of radial location in the disc. Here we interpret the results in terms of the water ice line at around 3 AU from the star, using a water ice mass fraction of $0.571\%$ \citep{lodders2003}.

Condensation is considered as a neighbour interaction. A linear scaling between number of particles and number of calculations is obtained by use of a mapping scheme, where the simulation domain is divided into a number of grid cells, with particle interactions possible between particles within the same grid cell only. 

Initially, both vapour and ice particles have a Gaussian distribution in the vertical direction, following the gas distribution. At the beginning of a simulation particles are set to be ice or vapour, depending on whether they are located within or outside of the condensation region.

\subsection{Particle sizes}

Particles are coupled to the gas via drag forces and can be characterized by their dimensionless friction time \citep{weidenschilling1977}, or Stokes number,
\begin{align}
&\varOmega\tau_{\rm f} ^{\rm (Ep)}=\frac{R\rho_\bullet}{H\rho_{\rm g}}&{\rm for}\,\,R<(9/4)\lambda\,,&\label{eq:epstein}\\ 
&\varOmega\tau_{\rm f} ^{\rm (St)}=\frac{4}{9}\frac{R^2\rho_\bullet}{H\rho_{\rm g}\lambda}&{\rm for}\,\,R>(9/4)\lambda\,,\label{eq:stokes}&
\end{align}
where $\lambda$ is the mean free path of the gas and $\rho_\bullet$ and $\rho_{\rm g}$ are the material and gas density. The superscripts $\rm (Ep)$ and $\rm (St)$ denote Epstein and Stokes drag regime, the two drag regimes relevant for the particle sizes and gas density considered in this paper. We formulate the particle radius $R$ in units of $\rm R_1$, where $R/R_1=\varOmega\tau_{\rm f}^{\rm (Ep)}$ so that 
\begin{align}
R_1=\frac{H\rho_{\rm g}}{\rho_\bullet}\,.
\end{align}
To find the corresponding particle size in meters, we use the material density of ice $\rho_\bullet\approx1\,\rm g\,cm^{-3}$, the midplane gas density $\rho_{\rm g}$ and column density $\Sigma$ from the Minimum Mass Solar Nebula (MMSN) model of \citet{hayashi1981},
\begin{align}
&\rho_{\rm g}=\frac{\varSigma(r)}{\sqrt{2\pi}H(r)}\,,\\
&\varSigma=1700\,{\rm g\,cm^{-2}}\left( \frac{r}{\rm AU} \right)^{-1.5}\,,\label{eq:cd}
\end{align}
giving the scaling of 
\begin{align}
R_1\approx1.3\,{\rm m}\left(\frac{r}{3\, \rm AU}\right)^{-1.5}
\end{align}
close to the water ice line.

\subsection{Condensation scheme}

The rate of change in particle mass due to condensation and sublimation is
\begin{align}
\frac{{\rm d}m}{{\rm d}t} = 4\pi R^2v_{\rm th}\rho_{\rm v} \left(1-\frac{P_{\rm sat}}{P_{\rm v}}\right) \,
\label{eq:cond}
\end{align}
\citep{supulverlin2000}. Here, $v_{\rm th}$ is the thermal velocity of vapour, $\rho_{\rm v}$ is the vapour density and $P_{\rm sat}$ and $P_{\rm v}$ is the saturated vapour pressure and vapour pressure, respectively. Both the vapour pressure and the saturated vapour pressure can be expressed as densities using the ideal gas law. The vapour pressure can be written as
\begin{align}
P_{\rm v}=\frac{k_{\rm  B}T}{m_{\rm v}}\rho_{\rm v}\,,
\end{align}
and the saturated vapour pressure can be expressed in an equivalent way. Here, $k_{\rm B}$ is the Boltzmann constant, $T$ is the temperature, $\rho_{\rm v}$ is the vapour density and $m_{\rm v}$ is the mass of one vapour particle. Assuming spherical particles Eq.\ \ref{eq:cond} can be rewritten in terms of particle radius $R$ and using vapour densities instead of pressures,
\begin{align}
\frac{{\rm d}R}{{\rm d}t} = \frac{v_{\rm th}}{\rho_\bullet}\left(\rho_{\rm v}-\rho_{\rm sat}\right) \,.
\end{align}
In the limit where $\rho_{\rm v}\ll\rho_{\rm sat}$ sublimation dominates the time evolution. The sublimation time scale in terms of particle radius is then
\begin{align}
\tau_{\rm s,\, R} & = \frac{R \rho_\bullet}{v_{\rm th} \rho_{\rm sat}}\,,
\end{align}
and in terms of mass 
\begin{align}
\tau_{\rm s} & = \frac{1}{3}\frac{R \rho_\bullet}{v_{\rm th} \rho_{\rm sat}}\,.
\end{align}
Correspondingly, when $\rho_{\rm sat}\ll\rho_{\rm v}$ condensation dominates, with condensation time scale in terms of radius
\begin{align}
\tau_{\rm c,\,R} & = \frac{R \rho_\bullet}{v_{\rm th}\rho_{\rm v}}\,,
\end{align}
and in terms of mass
\begin{align}
\tau_{\rm c} & = \frac{1}{3}\frac{R \rho_\bullet}{v_{\rm th}\rho_{\rm v}}\,.
\label{eq:condtimescale}
\end{align}
In equilibrium $\rho_{\rm v}=\rho_{\rm sat}$ with $\tau_{\rm s}=\tau_{\rm c}$. By definition $\rho_{\rm v}=f_{\rm H_2O}\rho_{\rm g}$, where $f_{\rm H_2O}$ is the water mass fraction, at an ice line. The rapid increase of the temperature into the condensation region implies $\rho_{\rm sat}\gg \rho_{\rm g}$ and hence $\tau_{\rm s}\ll \tau_{\rm f}$, allowing for modelling of sublimation as an instantaneous process as long as the particle friction time, $\tau_{\rm f}$, is shorter than the characteristic time-scale of the turbulent eddies, $\varOmega^{-1}$. 

Condensation can not be assumed to occur instantaneously, as $\rho_{\rm v}$ never increases much beyond $f_{\rm H_2O}\rho_{\rm g}$. The condensation time scale is proportional to the size of the particle, with the dimensionless condensation time scale being
\begin{align}
\varOmega\tau_{\rm c}=\frac{\Omega\tau_{\rm f}^{\rm (Ep)}}{3}\frac{\rho_{\rm g}}{\rho_{\rm v}}\,,
\label{eq:dimlesscondtimescale}
\end{align}
assuming $c_{\rm s}\approx \sqrt{\frac{\mu_{\rm H_20}}{\mu_{\rm H_2}}}v_{\rm th}$, where $\mu_{\rm H_20}$ and $\mu_{\rm H_2}$ are the mean molecular weights of water and molecular hydrogen. The condensation process is modelled by a Monte Carlo approach, where the probability of condensation for vapour onto an ice particle in the condensation zone is proportional to the size of the available ice surface. This is further described in Section \ref{sec:cond}.

\subsection{Superparticle approach}

\begin{figure}
\includegraphics[width=8.8cm]{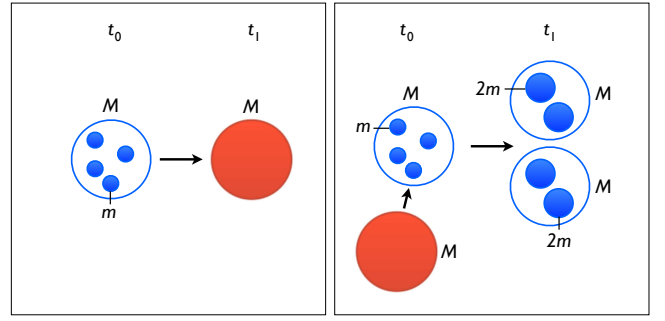}
\caption{Sketch of sublimation and condensation for superparticles. Blue represents ice and red vapour. Left panel: Ice superparticle, representing a mass $M$ in the form of 4 physical ice particles, each of mass $m$, sublimates and turns into a vapour superparticle of mass $M$. Right panel: Vapour superparticle of mass $M$ condenses onto an ice superparticle of mass $M$, representing the mass $4\, m$. The physical ice particles are after the event represented by two ice superparticles, each representing a total mass $M$, but now in the form of 2 physical ice particles each with mass $2\,m$. Both mass and the number of superparticles are conserved between $t_0$ and $t_1$. In the case of condensation, also the number of physical particles is conserved.}
\label{fig:superparticles}
\end{figure}

The ice and vapour components are modelled using a superparticle approach, related to the coagulation algorithm of \citet{zsomdullemond2008}, where a superparticle is a numerical representation of a large number of physical particles with identical properties. Here the most important properties are the physical state (ice or vapour), size  of constituent particles if ice, and material density. The number of superparticles $N$ is much smaller than the number of physical particles, so that superparticle $i$ represents $n_i$ physical particles. In these simulations, typically $N=10\,000$ for a simulation area of $12\, H$ in the vertical direction and $1.5\,H$ in the radial direction, a number chosen to be computationally easy to handle. The number of superparticles is fixed throughout a simulation, and each superparticle represents an equal and fixed mass $M$ that does not change. The mass $m$ of the physical particles represented by the superparticle does however change, and so does $n_i$, the number of physical particles represented by a superparticle, as the total mass represented by the superparticle is always $M=m_i\,n_i$. Also when an ice particle undergoes a phase transition and becomes vapour the total mass $M$ of the superparticle stays constant. The superparticle approach to condensation and sublimation is sketched in Fig.\ \ref{fig:superparticles}.

\subsection{Random walk of particles}
\label{sec:randomwalk}

In this section the modelling of particle motions is described. We start by introducing the random walk used to describe the turbulent motions of the gas, sufficient for modelling small particle motions, and then move on to include the more complex behaviour of larger particles. As particles grow larger, additional effects need to be taken into account. Firstly, the random step length becomes smaller, as larger particles follow the turbulent eddies a shorter length each time step. Secondly, for larger particles gravity towards the midplane has an increasingly important effect, causing particles to sediment towards the midplane. Finally, radial drift towards the central star must be taken into account.

With the dimensionless constant $\alpha$ introduced by \citet{shakurasunyaev1973}, the turbulent diffusion coefficient of gas and small particles in an accretion disc can be written as
\begin{equation}
D=\alpha c_{\rm s} H\,.
\label{eq:alphaprescription}
\end{equation}
We here assume that particles move with velocity $v\sim\sqrt{\alpha}c_{\rm s}$, and that the length a particle moves coherently is $l\sim\sqrt{\alpha}H$, which gives rise to $D\sim vl\sim\alpha c_{\rm s}H$ by mixing length arguments. The correlation time is defined as
\begin{equation}
\tau=\frac{l}{v}=\varOmega^{-1}\,.
\end{equation}
This means that the effect of turbulence on a small particle is to move it  in a random direction with a characteristic speed $v$, during the correlation time $\tau$. The characteristic length-scale, velocity-scale and time-scale are all set by the turbulence. We use a default $\alpha$-value of $10^{-2}$, motivated by observations of accretion rates in young stars \citep{hartmannetal1998}. However, the prescence of a dead zone can give a lower $\alpha$-value \citep{flemingstone2003, oishimaclow2009}, and there are also observations suggesting $\alpha=10^{-4}$ in protoplanetary discs \citep{muldersdominik2012}. Therefore we also run simulations with $\alpha=10^{-3}$ and $\alpha=10^{-4}$.

The distance a particle moves each time step, where one time step is $\Delta t=\varOmega^{-1}$, is found by equating the generic expression for the diffusion coefficient of a random walk in 2D, $D=l^2/(4\tau$), with that of diffusion described by Eq.\ \ref{eq:alphaprescription}, giving
\begin{equation}
l=2\sqrt{\alpha}H\,.
\label{eq:steplength}
\end{equation} 
Practically, the random walk for particles coupled to the gas is modelled by setting the length the particle moves in the radial direction $\Delta r$ and in the vertical direction $\Delta z$ to
\begin{align}
 (\Delta r)_{\rm rw}&=l\,\cos\theta \, ,\label{eq:dyn_lr}\\
 (\Delta z)_{\rm rw}&=l\,\sin\theta\,.
\label{eq:dyn_l}
\end{align}
Here $0\leq\theta< 2\pi$ is chosen randomly for each particle and time step, giving a random direction, but a fixed length, for how a particle moves in one time step. This forms the basis of the particle dynamics used in this code, and is valid for small particles perfectly coupled to the turbulent gas. However, for larger particles this simple approximation breaks down. A number of corrections to the random walk model will now be introduced, one at a time, before arriving at the final model for the particle dynamics.

Firstly, the random step length must be adjusted, as larger particles move shorter distances with the turbulent eddies each time step. The diffusion coefficient $D= vl= l^2/\tau= l^2\Omega$ is modified to 
\begin{equation}
D^*=\frac{D}{1+(\varOmega\tau_{\rm f})^2}\,
\end{equation}
following the analytical theory developed and tested by \citet{youdinlithwick2007}. This gives a modified size-dependent step length of 
\begin{equation}
l^*=\frac{2\sqrt{\alpha}H}{\sqrt{1+(\varOmega\tau_{\rm f})^2}}=\frac{l}{\sqrt{1+(\varOmega\tau_{\rm f})^2}}\,.
\end{equation}
For small particles $l^*\approx l$, whereas for larger particles $l^*<l$. Separating the radial and vertical directions, Eqs.\ \ref{eq:dyn_lr} and \ref{eq:dyn_l} are replaced by
\begin{align}
(\Delta r)_{\rm rw}&=\frac{l\,\cos\theta}{\sqrt{1+(\varOmega\tau_{\rm f})^2}}\,, \\
 (\Delta z)_{\rm rw}&=\frac{l\,\sin\theta}{\sqrt{1+(\varOmega\tau_{\rm f})^2}}\,.
\label{eq:dyn_l2}
\end{align}
As particles grow larger, and drag forces decrease in strength, the gravity towards the midplane influences the particle motion more, so that large particles sediment to the midplane. Since this affects the vertical direction only, it is only necessary to modify $\Delta z$. This is done by changing the step size in the vertical direction to $(\Delta z)_{\rm tot}=(\Delta z)_{\rm rw}+(\Delta z)_{\rm sed}$, where $(\Delta z)_{\rm sed}$ is an increasing function of particle size. The sedimentation step length is set by the terminal velocity of the particles $v_{\rm t}=-\tau_{\rm f}\Omega^2z$. However, in order to modify the step length correctly, we must ensure that the resulting particle scale height is the same as the analytically expected scale height. The analytically expected particle scale height, for a given particle size and turbulence strength, resulting from diffusion and sedimentation is
\begin{equation}
\frac{H_{\rm p}}{H}=\sqrt{\frac{\alpha}{\varOmega\tau_{\rm f}+\alpha}}\,,
\label{eq:randomwalk_hphg}
\end{equation}
as determined in computer simulations of particles in turbulent protoplanetary discs \citep{carballidoetal2006} and explained in the analytical diffusion framework of \citet{youdinlithwick2007}. This can now be compared with the particle scale height arising from considering a combination of a random walk and terminal velocity sedimentation, which is what is used for modelling. The random walk of particles mimics turbulent diffusion with coefficient $D=\alpha c_{\rm s}H$, and sedimentation-diffusion equilibrium occurs at particle scale height
\begin{equation}
\frac{H_{\rm p}}{H}=\sqrt{\frac{\alpha}{\varOmega\tau_{\rm f}}}\,.
\label{eq:randomwalk_hphgwrong}
\end{equation}
To construct a model that gives the correct scale height, as given by Eq.\ \ref{eq:randomwalk_hphg}, 
we therefore need to modify the dimensionless friction time in the terminal velocity expression to 
\begin{equation}
(\varOmega\tau_{\rm f})^*=\varOmega\tau_{\rm f}+\alpha\,.
\end{equation}
Effectively we let even tiny particles sediment in order to follow the Gaussian distribution of the gas, rather than the uniform distribution resulting from a pure random walk. Changing $\varOmega\tau_{\rm f}\;\rightarrow\;(\varOmega\tau_{\rm f})^*$ in Eq.\ \ref{eq:randomwalk_hphgwrong} reproduces the correct particle scale height in Eq.\ \ref{eq:randomwalk_hphg}. With the modified dimensionless friction time $(\varOmega\tau_{\rm f})^*$ in the terminal velocity expression, the step size in the vertical direction can be written as
\begin{equation}
( \Delta z)_{\rm tot}=( \Delta z)_{\rm rw}-(\varOmega\tau_{\rm f}+\alpha)\varOmega z \Delta t\,.
\label{eq:dyn_sed1}
\end{equation}
However, this expression leads to a numerical instability in which large particles overshoot the midplane as they sediment, since for $\varOmega\tau_{\rm f}>1$ and $\Delta t=\varOmega^{-1}$ we get
\begin{equation}
(\varOmega\tau_{\rm f})^*\varOmega z \Delta t>z\,.
\end{equation}
This gives particle oscillations that are amplified in time. To eliminate this problem, Eq.\ \ref{eq:dyn_sed1} is modified to
\begin{equation}
( \Delta z)_{\rm tot}= (\Delta z)_{\rm rw}-\frac{(\varOmega\tau_{\rm f}+\alpha)\varOmega z \Delta t}{1+(\varOmega\tau_{\rm f})^2}\,,
\label{eq:dyn_sed2}
\end{equation}
following the scheme of \citet{youdinlithwick2007}, which eliminates the amplifying of the vertical particle oscillations. Instead it gives larger particles a longer settling time, corresponding to the time they would have spent oscillating before settling. 
For a comparison between the particle scale height resulting from Eq.\ \ref{eq:dyn_sed2} and the theoretical scale height from Eq.\ \ref{eq:randomwalk_hphg}, see Fig.\ \ref{fig:dynamicstest}. The correlation between the analytical and modelled curve in this figure shows that the particle dynamics in the vertical direction are correctly modelled by Eq.\ \ref{eq:dyn_sed2}.

The step in the radial direction needs to be modified due to the radial drift towards the central star. This is also particle size dependent, and $(\Delta r)_{\rm tot}$ is modified to
\begin{equation}
(\Delta r)_{\rm tot}=(\Delta r)_{\rm rw}+(\Delta r)_{\rm rd}=(\Delta r)_{\rm rw}-\frac{2\eta v_{\rm K}}{\varOmega\tau_{\rm f} + (\varOmega\tau_{\rm f})^{-1}}\Delta t\,,
\label{eq:radialdrifteq}
\end{equation}
where $v_{\rm K}$ is the Keplerian velocity and $\eta v_{\rm K}$ is the velocity difference between the gas and the particles, following \citet{weidenschilling1977}. This gives a radial drift velocity that peaks for $\varOmega\tau_{\rm f}=1$ particles. We use $\eta v_{\rm K}=0.05 c_{\rm s}$, a reasonable value at $3\,\rm AU$ in the MMSN \citep{cuzzietal1993, chiangyoudin2010}.

The friction time of a particle is inversely proportional to the gas density, $\varOmega\tau_{\rm f}\propto\rho_{\rm g}^{-1}$, which decreases in the vertical direction with distance from the midplane. Taking into account the Gaussian distribution of the gas around the midplane, the dimensionless friction time of a particle is modified according to
\begin{align}
\varOmega\tau_{\rm f}\rightarrow \varOmega\tau_{\rm f}  \epsilon\, ,
\end{align}
where 
\begin{align}
\epsilon={\rm exp}\left[z^2/(2H^2)\right]\,.
\end{align}
Simulations have shown that the heterogeneous vertical gas distribution leads to a turbulent $\alpha$-value increasing with height above the midplane \citep{fromangnelson2009}. We do however not take this effect into account in our simulations. 

For large simulation domains radial gas density gradients are also of importance, and the random walk would then have to be adjusted in order to mimic the motion of especially small particles, well coupled to the gas \citep{hughesarmitage2010}. Since the model used in this paper is local, we ignore this effect.

The equations used to describe particle dynamics in the code are thus
\begin{align}
\Delta z&=\frac{l\,\sin\theta}{\sqrt{1+(\varOmega\tau_{\rm f}\epsilon)^2}}
-\frac{(\varOmega\tau_{\rm f}\epsilon+\alpha)\varOmega z}{1+(\varOmega\tau_{\rm f}\epsilon)^2}\Delta t\,,
\label{eq:deltaz}\\
\Delta r&=\frac{l\,\cos\theta}{\sqrt{1+(\varOmega\tau_{\rm f}\epsilon)^2}}-\frac{2\eta v_{\rm K}}{\epsilon\varOmega\tau_{\rm f} + (\epsilon\varOmega\tau_{\rm f})^{-1}}\Delta t\,,\label{eq:deltar}
\end{align}
where the first term for both the radial and vertical direction describe the random walk, with a step length modified according to particle size, and the second term describes sedimentation and radial drift, for the vertical and radial direction, respectively.

\subsection{Transition between drag regimes}

For small particles with a radius of less than (9/4) of the gas mean free path, Epstein drag applies, whereas for larger particle Stokes drag is the relevant regime. The transition size given by Eqs.\ \ref{eq:epstein} and \ref{eq:stokes} can be rewritten in dimensionless friction time as
\begin{align}
\left(\varOmega\tau_{\rm f}\right)^{\rm (Ep)}<\frac{9}{4}\frac{\rho_\bullet}{\rho_{\rm g}}\frac{\lambda}{H}
=\frac{9\pi}{2}\frac{\rho_\bullet\mu H}{\varSigma^2\sigma_{\rm mol}}\approx0.5\,,
\end{align}
when applied to the midplane of the disc. The mean molecular weight is $\mu=3.9\times10^{-24}\,\rm g$, the molecular cross-section of molecular hydrogen is $\sigma_{\rm mol}=2\times10^{-15}\,\rm cm^2$ \citep{nakagawaetal1986} and the column density $\varSigma$ at 3 AU is given by Eq.\ \ref{eq:cd}. 

Converting dimensionless friction time in the Epstein drag regime to the Stokes drag regime is done using
\begin{align}
\varOmega\tau_{\rm f}^{\rm (St)}=\frac{2}{9\pi}\frac{\varSigma^2\sigma_{\rm mol}}{\rho_\bullet \mu H}\left(\varOmega\tau_{\rm f}^{\rm (Ep)}\right)^2
\approx2.0\left(\varOmega\tau_{\rm f}^{\rm (Ep)}\right)^2\,,
\end{align}
where the scale height is given by the MMSN model \citep{hayashi1981} as
\begin{align}
H=0.033\left(\frac{r}{\rm AU}\right)^{1.25}\,.
\end{align}

\section{Condensation in a Monte Carlo scheme}
\label{sec:cond}

We treat condensation as a neighbour interaction, and require vapour to be present near an ice particle for interaction to occur. In this model ``near'' means being in the same grid cell, so that it is possible for an ice particle to interact only with vapour within the same grid cell. Condensation in each time step is thus decided one grid cell at a time. Any grid cell can, and most often does, harbour more than one ice particle. Therefore, whether or not condensation takes place in a given grid cell and a given time step, is decided in a two-step process for each vapour particle present in the grid cell.

The two-step procedure is here described for one grid cell and one time step $\Delta t$, with one vapour particle present in this grid cell. In the case where more than one vapour particle is present in the grid cell, the two-step procedure is repeated for each vapour particle present. The first step is to decide whether this vapour particle condenses onto any of the ice particles. If it does, the second step is to decide which of the ice particles it condenses onto. In this scheme, any vapour particle can only condense onto one single ice particle, so it can not be shared amongst several particles. One ice particle can on the other hand experience several growth events during one time step, if several vapour particles are present.

As a first step it is decided whether or not the vapour particle is involved in a condensation event during the time step $\Delta t$. In order to do this the total interaction probability for all ice particles in the grid cell is needed. The interaction probability for ice particle $i$ is
\begin{equation}
p_i = 1-{\rm exp}(-\Delta t/\tau_i)\,,
\end{equation} 
where $\tau_i$ is the condensation time scale for particle $i$, given by Eq. \ref{eq:dimlesscondtimescale}. The expression for interaction probability is chosen to always give $0\leq p_i<1$, for any $\Delta t$, also for $\Delta t\gg \tau_i$. The algorithm proceeds by calculating the probability that no interaction occurs. For $n$ ice particles, the total probability that no interaction occurs is
\begin{equation}
p_0 = (1-p_1)\cdot(1-p_2)\cdot \,\cdots \, \cdot(1-p_n)\,.
\end{equation}
To decide whether condensation occurs, a random number $r_1\in[0,1[$ is generated. If $r_1$ is smaller than $p_0$ nothing happens, whereas if $r_1$ is larger than $p_0$ vapour condenses onto one of the ice particles. 

If condensation occurs, a second step is needed in order to decide onto which of the ice particles in the grid cell. In this step it is no longer the absolute probability that is of interest, but the relative probabilities of the ice particles in the grid cell. This can be expressed as
\begin{equation}
p^*_i=\frac {\Delta t}{\tau_i}\,.
\end{equation}
The relative probability for each of the ice particles in the grid cell are placed in a sequence, 
\begin{equation}
0,\,p^*_1,\,p^*_1+p^*_2,\,p^*_1+p^*_2+p^*_3,\,\dots\,,\sum_ip^*_i\,,
\end{equation}
and a new random number is generated, $r_2\in[0,\sum_ip^*_i[$ . Which interval in the sequence $r_2$ falls in decides which ice particle the vapour particle condenses onto, such that $r_2$ falling in the interval $]p^*_{i-1},p^*_{i}]$ implies condensation onto ice particle $i$.

The condensation time scale in Eq.\ \ref{eq:condtimescale} is proportional to the radius of the particle, so that smaller particles have a shorter, and larger particles a longer, condensation time scale. This corresponds to a larger condensation probability for smaller particles and vice versa, which may seem counter-intuitive, but is explained by the fact that the simulation handles superparticles, each representing the same mass $M$. Therefore a superparticle representing small physical particles represents a larger combined surface area than one representing large particles. Condensation is thus more likely to happen to a superparticle representing small particles. Looking at a single-physical-particle-basis, the result is correct averaged over many timesteps and particles. A condensation event means doubling the mass of the physical particles involved, implying that a small particle experiences many condensation events, but the absolute growth in radius each event is small, whereas a single large physical particle experiences few condensation events, but with a large growth in absolute radius each event.

\section{Test problems}
\label{sec:test}

In order to understand the results of the computer simulations and to make sure that the code is functioning correctly, two major tests of the algorithms used were made. Firstly, the dynamical behaviour of the code was tested without including particle growth. Secondly, the particle growth algorithm was tested, without taking spatial dimensions into account.

\subsection{Test of the dynamical behaviour of the particles}

\begin{figure}
\includegraphics[width=8.8cm]{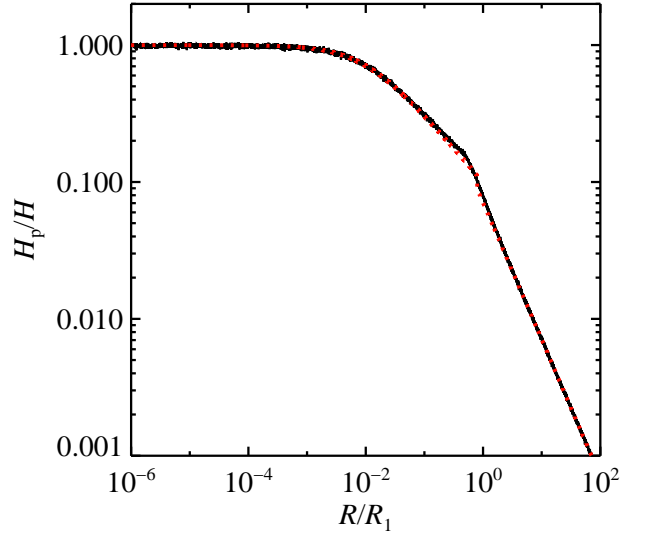}
\caption{Comparison between modelled and analytical particle scale height as a function of particle size, for a turbulence strength of $\alpha=10^{-2}$. The black full line denotes modelled values and the red dotted line analytical values. Small particles $R\lesssim10^{-3}$ are perfectly coupled to the gas, whereas larger particles sediment towards the midplane. The slope change at $R\approx0.5$ is due to the change from Epstein to Stokes drag regime. The particle size is given in units of $ R_1$, where $ R_1\approx1 \,{\rm m}$ at $r=3 \,{\rm AU}$. Overall, modelled and analytical values are in good agreement. }
\label{fig:dynamicstest}
\end{figure}

The dynamical behaviour of the particles is tested by excluding condensation and sublimation from the simulation. This means that particles move due to turbulence, stirring them up in a turbulent diffusion, and due to gravity directed towards the midplane, but no particle growth is included. We also ignore radial drift. The particles settle to an equilibrium, where the particle scale height depends on the size of the particles, since the strength of the coupling to the turbulent gas is a function of particle size. Fig.\ \ref{fig:dynamicstest} shows the particle scale height relative to the gas scale height $H_{\rm p}/H$ as a function of particle size $R$. The particles settle into a Gaussian distribution around the midplane, so the particle scale height in equilibrium can be retrieved as the root mean square of the particle positions,
\begin{equation}
\left(\frac{H_{\rm p}}{H}\right)_{\rm mod}=\sqrt{\frac{1}{N}\left( z_1^2+z_2^2+ \ldots +z_n^2 \right)}\,,
\end{equation}
where $n$ is the number of particles and $z_i$ is the vertical position of particles. This is compared to the theoretically expected particle scale height, given by an equilibrium between sedimentation and turbulent diffusion,
\begin{equation}
\left( \frac{H_{\rm p}}{H} \right)_{\rm theo}=\sqrt{\frac{\alpha}{\varOmega \tau _f +\alpha}}\,.
\end{equation}

Fig.\ \ref{fig:dynamicstest} shows how the particle scale height depends on the size of the particles, with the modelled values represented by the black full line and the expected analytical values as a red dotted line. The modelled line clearly follows the analytical curve, showing that the dynamical behaviour of the particles is as expected. The particle size is given in units of $R_1$, where $ R_1 \approx1 \,{\rm m}$ at $r=3 \,{\rm AU}$. From the figure one can see that small particles are fully coupled to the gas, as $H_{\rm p}/H\approx1$ for particles of $R \lesssim10^{-3} \,R_1$, corresponding  to mm-sized ice particles near the water ice line. Further, a slight change of the slope can be seen at $R\approx0.5 \,R_1$. This is caused by the change from Epstein to Stokes drag regime, which makes the particles decouple more quickly from the gas and hence increases sedimentation.  

\subsection{Test of the condensation algorithm}
We test the condensation algorithm by letting particles grow via condensation, without including the spatial dimensions. The ice line is therefore not included, so sublimation is not taken into account. The number of ice superparticles is known, as is the number of vapour superparticles. The total mass available for growth is known, since we run the simulation until all vapour has condensed onto the ice particles. By tracking the initial and final mass of all superparticles the growth in radius is followed, and can be compared to a theoretical expectation.

\begin{figure}
\includegraphics[width=8.8cm]{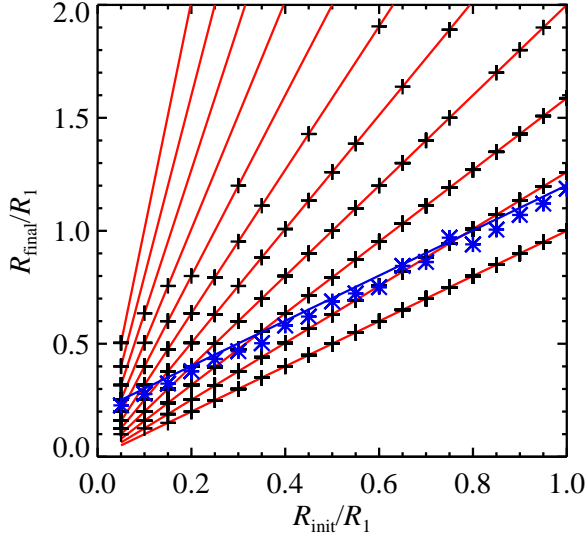}
\caption{Comparison between average modelled growth and analytically expected growth of ice particles. The final particle size as a function of initial particle size is shown for 2000 ice superparticles distributed evenly in 20 different initial size bins over the range $R/R_1=[0,1]$ , and with 20\,000 vapour superparticles added. Black crosses show modelled values and blue stars show the average growth in each size bin. The modelled average size roughly agrees with the analytical value, shown as a blue line. Red lines denote expected size for a number of condensation events between 0 (lower line) and 10 (upper line).}
\label{fig:condtest_nonrandom}
\end{figure}

In reality, condensation is a continuous process on macroscopic scales, where single water molecules condense onto ice particles so that growth happens little by little all the time. In the Monte Carlo scheme used in this code this continuous behaviour is modelled by a discrete growth process, where a condensation event involves one ice superparticle and one vapour superparticle. The mass of the physical ice particles represented by the ice superparticle doubles as the vapour superparticle completely condenses onto the ice, implying that there is no vapour superparticle left after the event. To conserve the number of ice particles and superparticles (for easier coding purposes) this implies that the physical ice particles that before the event were represented by the one ice superparticle involved in the event, now are represented by two superparticles, i.e. the number of physical ice particles before the event are after the event split between two ice superparticles. The mass of the physical particles involved in an event thus doubles as a condensation event occurs. As particle radius is connected to mass as $m\propto R^3$, for any number of condensation events $n_{\rm ce}$ the radius increases as
\begin{equation}
R\rightarrow \left( 2^{ n_{\rm  ce}} \right)^{1/3}R\,.
\end{equation}
The final radius $R_{\rm final}$ is plotted as a function of $R_{\rm init}$ as black crosses in Fig.\ \ref{fig:condtest_nonrandom}. In this test 2000 ice superparticles were distributed evenly in 20 different initial size bins over the range $R/R_1=[0,1]$ , and 20\,000 vapour superparticles were added. The possible growth in radius from mass doubling events is plotted as red lines in Fig.\ \ref{fig:condtest_nonrandom}, for a different number of condensation events $0\le {\rm n_{ce}}\le10$. From the figure it is clear that all modelled values indeed fall on these lines, and never in-between, as expected because of the discrete nature of the Monte Carlo method used. 

The modelled particle growth can be compared to the analytically expected $\Delta R$. For a physical ice particle $i$ the total mass growth due to condensation during the time $\Delta t$ is 
\begin{equation}
\Delta m_i=m_i-m_{0,i}=\frac{4\pi}{3}\left( R_i^3-R_{0,i}^3 \right)\rho_\bullet\,,
\end{equation}
where $\rho_\bullet$ is the material density of ice, $R_{0,i}$ is the initial and $R_i$ the final radius of the ice particle. However, the physical particles are represented by superparticles in the code, and the mass growth must therefore be given for superparticles and not physical particles. One superparticle represents the mass $M_i=n_im_i$, where $n_i$ is the number of physical particles represented by the superparticles. 

The total mass of vapour that condenses onto the ice particles is given by the difference between the sum of final and initial ice superparticles, 
\begin{equation}
M_{\rm v}=\rho_{\rm v}V=\frac{4\pi}{3}\rho_\bullet\left( \sum\limits _i  n_i R_i^3 - \sum\limits _in_{0,i}R_{0,i}^3\right) \,.
\label{eq:condtest_spsum}
\end{equation}
The total number of physical ice particles is constant, as is the total number of ice and vapour superparticles, but the number of ice superparticles is not. As the physical particles represented by one superparticle grow in mass, the number of physical particles represented by the superparticle decreases since the total mass of a superparticle is constant. In order to conserve the number of physical particles, total number of superparticles and the total mass of each superparticle, the number of ice superparticles increases as the physical particles grow in mass, so that each superparticle represents fewer and fewer physical particles, which is why the two sums in Eq.\ \ref{eq:condtest_spsum} are over different numbers of superparticles. 

The right handside of Eq. \ref{eq:condtest_spsum} can be rewritten using 
\begin{equation}
n_i=\frac{M_i}{m_i}=\frac{\rho_i V}{\rho_\bullet \frac{4\pi}{3}R_i^3}\,.
\label{eq:condtest_nm}
\end{equation}
Here $V$ is the volume represented by a superparticle, which in this one-grid-cell test problem is equal to the total volume of the box. By cancelling terms and gathering the densities on the left handside Eq.\ \ref{eq:condtest_spsum} can be rewritten as
\begin{equation}
\frac{\rho_{\rm v}}{\rho_i}=N_{\rm p}-N_{\rm p,0}\,
\end{equation}
where $N_{\rm p}$ and $N_{\rm p,0}$ is the final and initial number of ice superparticles, respectively. All superparticles represent the same density, and therefore the ratio between the two densities corresponds to the ratio between the number of vapour and ice superparticles represented, so that 
\begin{equation}
\frac{\rho_{\rm v}}{\rho_i}=\frac{N_{\rm v}}{1}\,,
\end{equation}
with $N_{\rm v}$ being the number of vapour superparticles added to the simulation. Eq.\ \ref{eq:condtest_spsum} therefore is equivalent to the very simple expression
\begin{equation}
N_v=N_{\rm p}-N_{\rm p,0}\,,
\end{equation}
which the code can be checked against, as both the number of vapour superparticles put into the system, and the initial number of ice superparticles, is specified, and the final number of superparticles is given as an output from the program. This analysis shows that the code is mass-conserving, by construction.

\begin{table*}
\caption{List of simulations, with references to relevant figures in the second column. The dimensions of the simulation box are given as $r_{\rm min}$, $r_{\rm max}$, $z_{\rm min}$ and $z_{\rm max}$ in gas scale heights $H$. The number of grid cells in the radial and vertical direction is given as $n_{ r} \times n_{z}$ and the number of particles as $n_{\rm p}$. Position of ice lines is given as $z_{\rm ice}/H$ for the atmospheric ice line, and $r_{\rm ice}/H$ for the radial ice line when applicable. Turbulence strength is given by $\alpha$ and the initial particle size $R_{\rm init}$ is given in units of $R_1$, approximately equivalent to size in meters.} 
\label{tab:sims}      
\centering                                      
\begin{tabular}{l l l l l l l l l l l l }          
\hline\hline                       
Run & Fig.\  & $r_{\rm min}/H$ & $r_{\rm max}/H$ & $z_{\rm min}/H$ & $z_{\rm max}/H$ & $n_{\rm r} \times n_{\rm z}$ & $n_{\rm p}$ & $z_{\rm ice}/H$ & $r_{\rm ice}/H$ & $\alpha$ & $R_{\rm init}/R_1$  \\    
\hline                                   
1a	& \ref{fig:zicedep} & $-0.75$ & 0.75 & $-6.0$ & $6.0$ & $64\times 512$ & $10^3,10^4, 10^5$ & $0.3$ & - & $10^{-2}$ & $10^{-3}$ \\
1b	& \ref{fig:zicedep} & $-0.75$ & 0.75 & $-6.0$ & $6.0$ & $64\times 512$ & $10^3,10^4, 10^5$ & $0.6$ & - & $10^{-2}$ & $10^{-3}$ \\
1c	& \ref{fig:zicedep} & $-0.75$ & 0.75 & $-6.0$ & $6.0$ & $64\times 512$ & $10^3,10^4, 10^5$ & $1.0$ & - & $10^{-2}$  & $10^{-3}$ \\
1d	& \ref{fig:zicedep} & $-0.75$ & 0.75 & $-6.0$ & $6.0$ & $64\times 512$ & $10^3,10^4, 10^5$ & $1.4$ & - & $10^{-2}$ & $10^{-3}$ \\
1e	& \ref{fig:zicedep} & $-0.75$ & 0.75 & $-6.0$ & $6.0$ & $64\times 512 $ & $10^3,10^4, 10^5$ & $1.8$ & - & $10^{-2}$ & $10^{-3}$ \\
1f	& \ref{fig:zicedep} & $-0.75$ & 0.75 & $-6.0$ & $6.0$ & $64\times 512$ & $10^3,10^4, 10^5$ & $3.0$ & - & $10^{-2}$ & $10^{-3}$ \\
\hline
2a	& \ref{fig:alphadep} & $-0.75$ & 0.75 & $-6.0$ & $6.0$ & $64 \times 512$ & $10^4$ & $[0.2,\, 3.0]$ & - & $10^{-4}$ & $10^{-4}$ \\
2b	& \ref{fig:alphadep} & $-0.75$ & 0.75 & $-6.0$ & $6.0$ & $64 \times 512$ & $10^4$ & $[0.2,\,3.0]$ & - & $10^{-3}$ & $10^{-4}$ \\
2c	& \ref{fig:alphadep} & $-0.75$ & 0.75 & $-6.0$ & $6.0$ & $64\times 512$ & $10^4$ & $[0.2,\,3.0]$ & - & $10^{-2}$ & $10^{-4}$ \\
\hline
3a	& \ref{fig:radialiceline} & $-0.75$ & 0.75 & $-6.0$ & $6.0$ & $64 \times 512$ & $10^4$ & $0.4$ & $-0.3$ & $10^{-2}$ & $10^{-3}$ \\
3b	& \ref{fig:radialiceline} & $-0.75$ & 0.75 & $-6.0$ & $6.0$ & $64\times 512$ & $10^4$ & $0.6$ & $-0.3$ & $10^{-2}$ & $10^{-3}$ \\
3c	& \ref{fig:radialiceline} & $-0.75$ & 0.75 & $-6.0$ & $6.0$ & $64\times 512$ & $10^4$ & $1.0$ & $-0.3$ & $10^{-2}$ & $10^{-3}$ \\
3d	&  \ref{fig:radialiceline} & $-0.75$ & 0.75 & $-6.0$ & $6.0$ & $64\times 512$ & $10^4$ &$1.4$ & $-0.3$ & $10^{-2}$ & $10^{-3}$ \\
3e	& \ref{fig:radialiceline} & $-0.75$ & 0.75 & $-6.0$ & $6.0$ & $64\times 512$ & $10^4$& $1.8$ & $-0.3$ & $10^{-2}$ & $10^{-3}$ \\
3f & \ref{fig:radialiceline} & $-0.75$ & 0.75 & $-6.0$ & $6.0$ & $64\times 512$ & $10^3, 10^4, 10^5$ & $3.0$ & $-0.3$ & $10^{-2}$ & $10^{-3}$ \\
\hline
4a	& \ref{fig:extbox_nopb} & $-0.75$ & 0.75 & $-6.0$ & $6.0$ & $64\times512$ & $10^4$ & $3.0$ & $-0.3$ & $10^{-4},10^{-3},10^{-2}  $ & $10^{-3}$ \\
4b	& \ref{fig:extbox_nopb} &  $-0.75$ & 2.25 & $-6.0$ & $6.0$ & $128\times 512$ & $2\times10^4$ & $3.0$ & $-0.3$ &  $10^{-4},10^{-3},10^{-2}$ & $10^{-3}$ \\
4c	& \ref{fig:extbox_nopb} &  $-0.75$ & 3.75 & $-6.0$ & $6.0$ & $192\times 512$ & $3\times10^4$ & $3.0$ & $-0.3$ & $10^{-4},10^{-3},10^{-2}$ & $10^{-3}$ \\
4d	& \ref{fig:extbox_nopb} & $-0.75$ & 5.25 & $-6.0$ & $6.0$ & $256\times 512$ & $4\times10^4$ & $3.0$ & $-0.3$ & $10^{-4},10^{-3},10^{-2}$ & $10^{-3}$ \\
\hline                                             
\end{tabular}
\label{tab:results}
\end{table*}

From Eq.\ \ref{eq:condtest_spsum} the theoretically expected $\Delta R$, equal for all particles, can be found. The final radius can be rewritten as $a_i=\Delta R+R_{0,i}$, where $R_{0,i}$ is the initial radius of the particle. The two sums are not necessarily over the same number of superparticles, as each condensation event introduces a new ice superparticle in the system, at the expense of the vapour particle representing the condensing vapour. The physical particles originally represented by the first ice superparticle are now split between the new and the old ice superparticle. The initial size of a physical particle represented by a converted vapour superparticle can therefore be found as the initial size represented by the first ice superparticle. Eq.\ \ref{eq:condtest_spsum} is thereby rewritten as
\begin{equation}
M_{\rm v}=\frac{4\pi}{3}\rho_\bullet\left[ \sum\limits_i n_i(\Delta R+R_{0,i})^3 - \sum\limits_{0,i} n_i R_{0,i}^3\right]\,,
\end{equation}
which, by using Eq.\ \ref{eq:condtest_nm}, cancelling terms and noting that (as in Eq.\ \ref{eq:condtest_spsum}) $\rho_{\rm v}/\rho_i=N_{\rm v}$, gives
\begin{equation}
N_{\rm v}=\sum\limits_i\frac{1}{R_i^3}(\Delta R + R_{0,i})^3-N_{\rm p,0}\,.
\end{equation}
The initial number of vapour and ice superparticles $N_v$  and $N_{\rm p,0}$ are known, as they are input parameters to the simulation. The radius represented by a superparticle in the end of the simulation, $a_{i}$, and when it is created, $R_{{\rm c},i}$, are both given as output from the code. The radius growth for each particle $\Delta R$ can therefore be found. For $N_v=20\,000$ and $N_{\rm p,0}=2000$, a value of $\Delta R \approx0.202$ is found. In Fig.\ \ref{fig:condtest_nonrandom} the blue full line represents the expected growth of the 2000 initial ice superparticles, $(R_{0,i}+\Delta a)$, as a function of initial radius $R_{0,i}$. Comparing the expected final particle radii with the actual final radii shows that for small particle sizes the Monte Carlo method works fairly well, whereas for larger particle sizes most particles experience too little (zero) growth and a few grow several orders of magnitude more than expected. This means that one must be careful not to draw conclusions from the growth of individual particles. However, statistically speaking, i.e.\ averaging over many particles, the result can be considered to be correct \citep{zsomdullemond2008}.

\begin{figure*}
{\includegraphics[width=18cm]{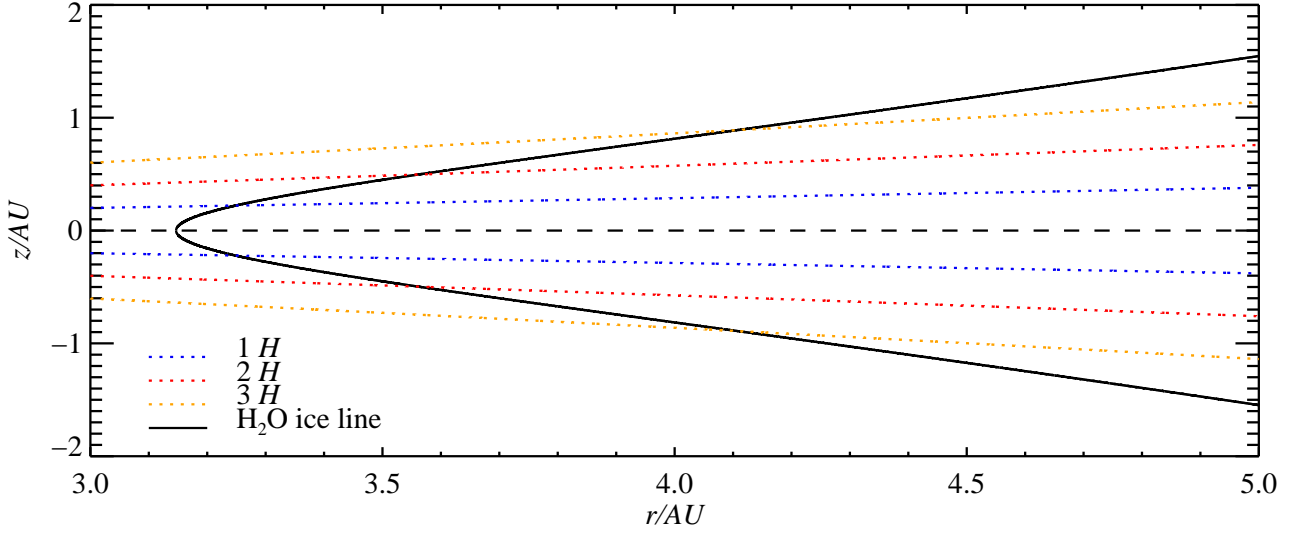}}
\caption{Pressure ice line for a vertically isothermal disc. The full black line denotes the ice line, whereas the coloured dotted lines show the location of $1\,H$, $2\,H$ and $3\,H$. As particles are preferably found within a few scale heights from the midplane, growth due to crossing of the atmospheric ice line is only important within less than $1\,\rm AU$ of the radial ice line.}
\label{fig:iceline}
\end{figure*}

\section{Results}
\label{sec:results}

The results are divided into two parts: runs including only the atmospheric ice line (Table \ref{tab:results}: runs 1 and 2) and  runs including both the radial and atmospheric ice lines (Table \ref{tab:results}: runs 3 and 4). We assess how the growth depends on turbulence strength in run 2 and 4, its dependence on atmospheric ice line position in runs 1, 2 and 3 and the dependence on box size in run 4. In all runs particle collisions are ignored, an assumption that we discuss in Section \ref{sec:collisions}.

\subsection{Atmospheric ice line}
\label{sec:atm}

\begin{figure*}
{\includegraphics[width=18cm]{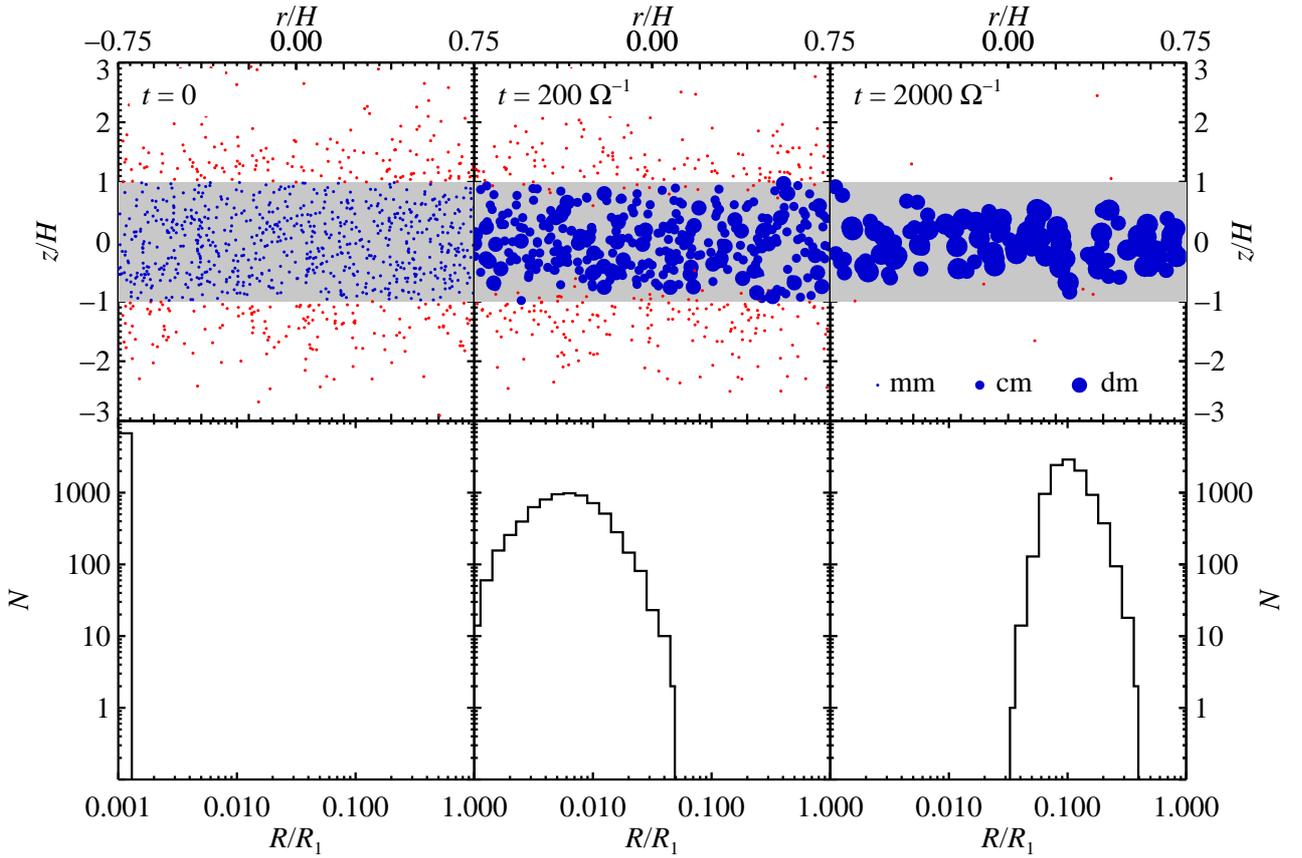}}
\caption{State of simulation with only the atmospheric ice line included, for $t=0,\,200\,\varOmega\,{\rm and}\, 2000\,\varOmega^{-1}$, from left to right. Upper panels: Distribution of blue ice particles and red vapour particles, with the grey area representing the condensation zone. The size of the blue dots are proportional to the size of the ice particles and the number of particles shown is inversely scaled with size for visibility. Lower panels: Evolution of the size distribution of ice particles at corresponding times. $N$ is the number of ice superparticles, and the size is given as $R/R_1$. At the water ice line, $R_1\approx1.3\,{\rm m}$.}
\label{fig:atm_timeevo}
\end{figure*}

We start by exploring the atmospheric ice line, not including the radial ice line. This means that the simulation domain is located just outside of the radial ice line, with a colder midplane and hotter outer layers \citep{dullemonddominik2004}. 

The thermal atmospheric ice line in a typical protoplanetary disc is located at $z\gtrsim3\,H$\citep{chianggoldreich1997}. However, the decrease of pressure with height above the midplane leads to a narrow radial zone just outside of the radial ice line where the atmospheric ice line is located at $z\le3\,H$. The pressure ice line in a vertically isothermal disc is shown in Fig.\ \ref{fig:iceline}. 

The height of the atmospheric ice line above the midplane $z_{\rm ice}$ is varied from $z_{\rm ice}=0.4\,H$ to $z_{\rm ice}=3\, H$. The lower limit is the lowest value for which not all ice particles immediately sublimate. A lower value gives a distance between the midplane and the ice line which is comparable to the length a particle moves during one time step (see Eq.\ \ref{eq:steplength}), making it possible for all ice particles to escape from the condensation region before any growth has taken place. Periodic boundary conditions are used in both the radial and vertical direction.

Fig.\ \ref{fig:atm_timeevo} shows the time evolution of a simulation with the atmospheric ice line at $z_{\rm ice}=1\,H$. The distribution of red vapour and blue ice particles is shown together with the size distribution of ice particles, for $t=0$, $t=200\,\varOmega^{-1}$ and $t=2000\,\varOmega^{-1}$. Vapour diffuse into the grey condensation region, condensing onto already existing ice particles. As small ice particles tend to diffuse out of the condensation region and sublimate, whereas larger ones stay in the midplane and experience continued growth, the result is a narrow size distribution with decimeter-sized ice pebbles residing in a midplane layer.

\subsubsection{Varying the atmospheric ice line position}
\label{sec:atm_1}

\begin{figure*}
{\includegraphics[width=18cm]{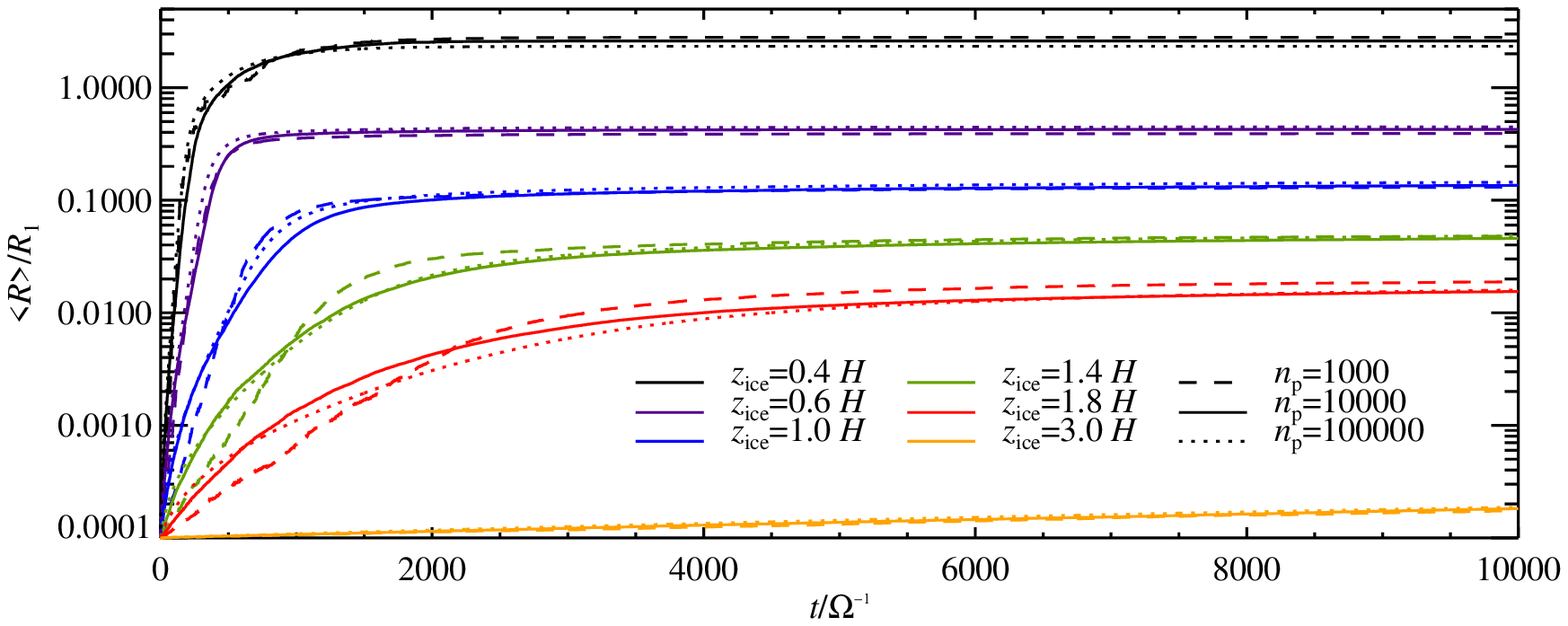}}
\caption{Mean ice particle size as a function of time for different positions of the atmospheric ice line. The different lines show the growth for different distances from the atmospheric ice line to the midplane, $z{\rm_{ice}}$. From top to bottom $z{\rm_{ice}}=\, 0.4\,H,\, 0.6\,H,\, 1.0\, H,\, 1.4\,H, 1.8\,H$ and $3.0 \,H$. Ice particles grow faster, and to larger sizes, the closer to the midplane the atmospheric ice line is. Full lines denote simulations with $10\,000$ particles, dashed lines $1000$ particles and dotted lines $100\,000$ particles.}
\label{fig:zicedep}
\end{figure*}

Simulations were run with the atmospheric ice line placed at different heights $z_{\rm ice}$ above the midplane. Decreasing $z_{\rm ice}$ corresponds in a physical protoplanetary disc to placing the simulation box radially closer in towards the radial ice line, where a hypothetical $z_{\rm ice}=0$ would correspond to the position of the radial ice line, and thus assessing the narrow radial zone where the pressure ice line dominates.

In Fig.\ \ref{fig:zicedep} the particle growth for different $z{\rm_{ice}}$ is shown. The mean ice particle size $\langle R\rangle$ in units of $\rm R_1$ is shown as a function of time in units of $\varOmega^{-1}$. The curves show, from top to bottom, $z_{\rm ice}=0.4\,H$ (black),  $z_{\rm ice}=0.6\,H$ (violet), $z_{\rm ice}=1.0\,H$ (blue), $z_{\rm ice}=1.4\,H$ (green), $z_{\rm ice}=1.8\,H$ (red) and $z_{\rm ice}=3.0\,H$ (yellow). As is clear from the figure, particles grow to larger sizes as the ice line is moved closer to the midplane. This is effectively due to the fact that for ice lines closer to the midplane, i.e.\ a narrower condensation region, vapour particles can easier penetrate to the larger ice particles residing in the midplane. 
Fig.\ \ref{fig:dynamicstest} shows how the particle scale height $H_{\rm p}$ decreases with increasing particle radius $R$. Small particles have a scale height comparable to the gas scale height $H_{\rm p, small}\approx  H$, whereas for larger particles $H_{\rm p, large}\ll  H$. Placing the ice line at a height larger than $1\, H$ is thus equivalent to saying that even the smallest particles tend to stay within the condensation zone. There is therefore very little material available for particle growth. The few particles that do move across the ice line and sublimate will mostly redistribute material among the particles at the border of the condensation zone. Growth in runs with $z_{\rm ice}\gtrsim 1\, H$ is both slow, and to modest particle sizes. For ice line positions of $z_{\rm ice}\geq3\,H$ the particle growth is on the order of $\Delta R\sim10^{-4}\,R_{\rm1}$ or less in $1000\,\varOmega^{-1}$. For the larger extent of the disc, where $z_{\rm ice}\gtrsim3H$, the atmospheric ice line is thus of little importance for particle growth by condensation.

In Fig.\ \ref{fig:zicedep} we show runs with 1000, 10 000 and 100 000 particles for each ice line position. The negligible difference between runs with 10 000 and 100 000 particles in combination with a much lower computational cost motivates the use of 10 000 particles as a reference resolution.

\subsubsection{Varying the turbulent $\alpha$-value}
\label{sec:atm_2}

\begin{figure*}
\includegraphics[width=18cm]{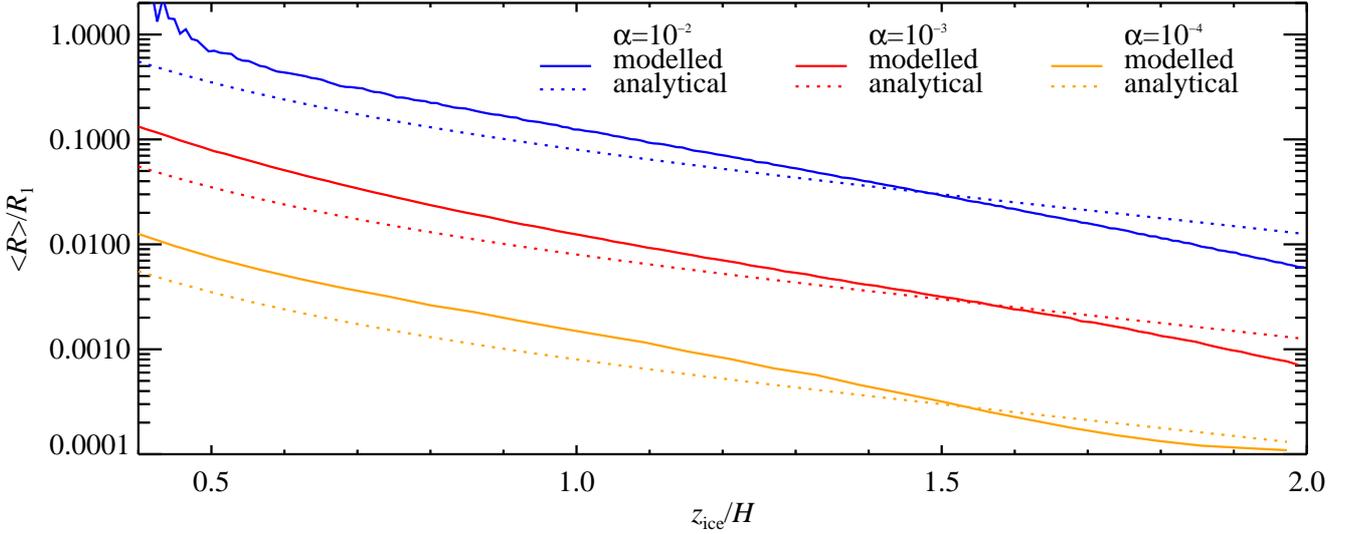}
\caption{Mean ice particle size as a function of the position of the atmospheric ice line and the strength of the turbulence. The mean particle size $\langle R \rangle$ is shown as a function of the distance from the atmospheric ice line to the midplane $z_{\rm ice}$, in gas scale heights $H$. The system is shown when the particle sizes have approximately reached an equilibrium, at $t=5000 \,\varOmega^{-1}$ for $\alpha=10^{-2}$, shown in blue, at $t=25\,000 \,\varOmega^{-1}$ for $\alpha=10^{-3}$, shown in red, and at $t=100\,000 \,\varOmega^{-1}$ for $\alpha=10^{-4}$, shown in yellow. The initial particle size is $10^{-4}\, R_1$, corresponding to approximately $0.1 \rm mm$ at the ice line. Solid lines are modelled values and dotted lines are analytical values, set by Eq.\ \ref{eq:hphgtheoline}. The modelled and analytical values follow the same slope in the applicable range.}
\label{fig:alphadep}
\end{figure*}

Simulations were run with different values of $\alpha$ in order to test the influence of turbulence strength on the results. In Fig.\ \ref{fig:alphadep} the mean particle size $\langle R\rangle$ is shown as a function of the height of the ice line above the midplane $z_{\rm ice}$, starting from an initial particle size of $10^{-4}\,\,R_1$. The system is shown when it has approximately reached an equilibrium, at $t=5000\,\varOmega^{-1}$ for $\alpha=10^{-2}$, shown in blue, at $t=25\,000\,\varOmega^{-1}$ for $\alpha=10^{-3}$, shown in red, and at $t=100\,000\,\varOmega^{-1}$ for $\alpha=10^{-4}$, shown in yellow. Full lines represent modelled values and dotted lines are analytical estimates. As can be seen, lowering the strength of turbulence gives less growth. This is to be expected as the main effect of decreasing the level of turbulence is to lower the effective particle scale height, set by an equilibrium between sedimentation and turbulent diffusion. As a comparison, dotted lines show the analytical relation between mean particle size and particle scale height in a system where vertical gravity and turbulent diffusion is dominating. The scale height of particles in terms of the gas scale height is
\begin{equation}
\frac{H_{\rm p}}{H}=\sqrt{\frac{\alpha}{\varOmega\tau_{\rm f}+\alpha}}\,.
\label{eq:hphgtheoline}
\end{equation}
Setting $H_{\rm p}\sim (1/c) \, z_{\rm ice}$ gives an analytical estimate of the maximum particle size achievable by condensation, shown in Fig.\ \ref{fig:alphadep}, as
\begin{equation}
\varOmega\tau_{\rm f}= \alpha\left\{\frac{1}{[ z_{\rm ice}/(c\,H)]^2}-1\right\}\,.
\end{equation}
We chose $c=3$, but even in this case the analytical expression underestimates the resulting $R$ slightly. Nevertheless our simulations show that particles grow approximately to a size where their scale-height is 1/3 of the ice line height.

\subsection{Atmospheric and radial ice line}

\begin{figure}
\includegraphics[width=8.8cm]{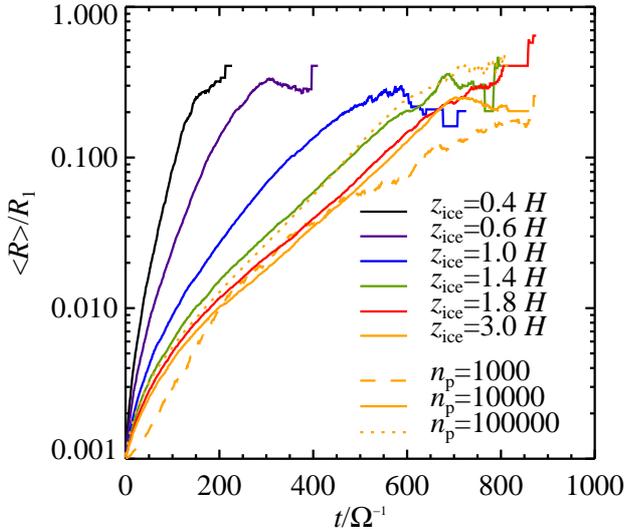}
\caption{Mean ice particle size $\langle R \rangle$ as a function of time in inverse Keplerian orbits $\varOmega^{-1}$ for simulations including both radial and atmospheric ice lines. The height of the atmospheric ice line above the midplane $z_{\rm ice}$ is denoted as: black 0.4 {\it H}, violet 0.6 {\it H}, blue 1.0 {\it H}, green 1.4 {\it H}, red 1.8 {\it H} and yellow  3.0 {\it H}. For the runs with $z_{\rm ice}=3.0\,H$, the dashed line corresponds to $1000$ particles and the dotted line to $100\,000$ particles. Full lines denote runs with $10\,000$ particles. The position of the radial ice line is $r_{\rm ice}=-0.3\,H$, and the strength of turbulence is $\alpha=10^{-2}$. }
\label{fig:radialiceline}
\end{figure}

Including both the atmospheric and radial ice lines is done by letting the simulation domain represent a region around the radial ice line. This means that ice particles can sublimate both by moving away from the midplane, and by moving closer to the central star. For a realistic atmospheric ice line position at $z_{\rm ice}\gtrsim3\,H$, sublimation by moving across the radial ice line is however much more important than by moving across the atmospheric ice line. The simulation domain is $12\,H$ in the vertical direction. In the radial direction the box size is varied from $1.5\,H$ to $5.25\,H$, with the number of particles and grid cells changed accordingly in order to keep the particle density and the effective resolution fixed. Periodic boundary conditions are used in the vertical direction, and reflective boundary conditions in the radial direction. In the vertical direction there is in effect no difference between $+z$ and $-z$, as the conditions above the midplane mirror those below it. In the radial direction, when including the radial ice line, there is no such symmetry. At $r=r_{\rm min}$ conditions such that ice sublimates prevail at all $z$, whereas at $r=r_{\rm max}$ the midplane is part of the condensation zone, where ice particles can exist without sublimating. Using periodic boundary conditions in the radial direction would therefore artificially introduce vapour in the system, causing a too large particle growth.

\subsubsection{Varying the atmospheric ice line position}
\label{sec:rad_1}

\begin{figure*}
{\includegraphics[width=18.0cm]{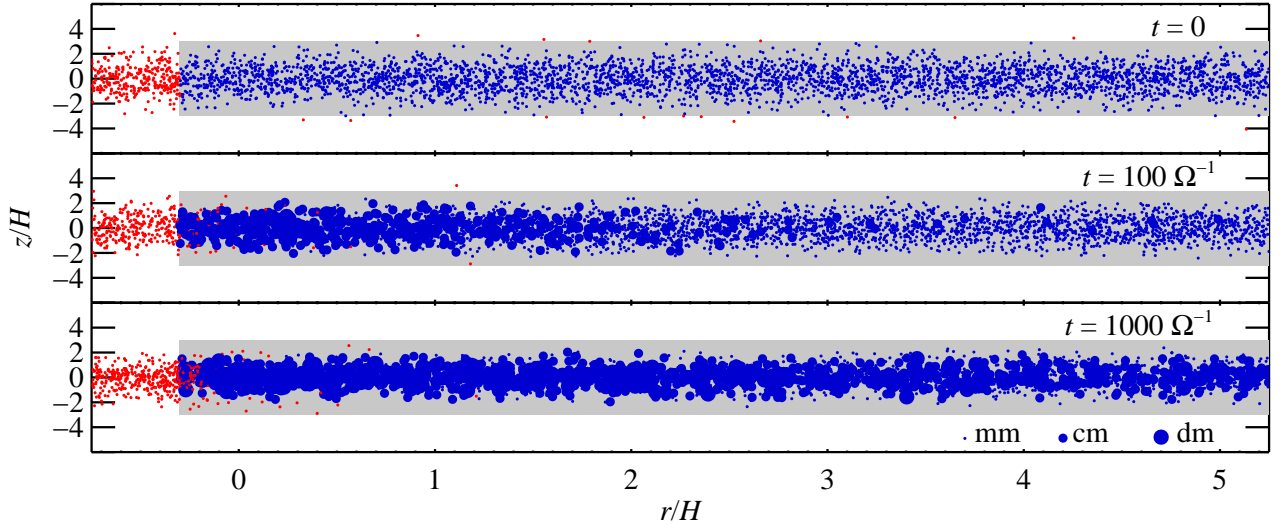}}
\caption{State of an extended simulation box with both the radial and the atmospheric ice line included, for $t=0,\,100\,{\rm and}\, 1000\,\varOmega^{-1}$, from top to bottom. The grey area represents the condensation zone, and ice and vapour is shown in blue and red, respectively. The sizes of the blue dots are proportional to the size of the ice particles and the number of particles shown is inversely scaled with size for visibility. The turbulent $\alpha$-value is $10^{-2}$.}
\label{fig:rad_timeevo}
\end{figure*}

Including the radial ice line greatly increases the growth efficiency. As shown in Sec.\ \ref{sec:atm}, models with only the atmospheric ice line gives negligible growth for ice line positions of $z_{\rm ice}\gtrsim3\,H$. However, taking also the radial ice line into account leads to particle growth beyond $0.1\,R_1$, corresponding to decimeter-sized pebbles at the ice line, within $1000\,\varOmega^{-1}$. 

Fig.\ \ref{fig:radialiceline} shows how the mean particle size $\langle R\rangle$ as a function of time varies with atmospheric ice line position $z_{\rm ice}$ when also the radial ice line is included. The colours show different heights of the atmospheric ice line above the midplane, with black denoting $ z_{\rm ice}=0.4\,H$, violet $ z_{\rm ice}=0.6\,H$, blue $ z_{\rm ice}=1.0\,H$, green $ z_{\rm ice}=1.4\,H$, red $ z_{\rm ice}=1.8\,H$ and yellow $z_{\rm ice}=3.0\,H$. Growth is somewhat faster for $z_{\rm ice}$ close to the midplane and slower for $ z_{\rm ice}$ further away from it. As previously discussed, this is due to the larger supply of material available for growth in a narrow condensation region, as compared to a wider condensation region. For ice line position above $3\,H$, a similar growth as for $z_{\rm ice}=3\,H$ is found.

Growth stops at $\langle R\rangle\approx R_1$ for all $z_{\rm ice}$. This is due to radial drift towards the star, which eventually removes all ice particles from the simulation. From Eq.\ \ref{eq:radialdrifteq} it can be seen that the radial drift inwards peaks for particles of $\langle R\rangle\approx R_1$. As particles reach this size they thus drift inwards, past the radial ice line, and sublimate.  

We also show the growth for ice line position of $z_{\rm ice}=3.0\,H$ for a smaller ($n_{\rm p}=1000$) and a larger ($n_{\rm p}=100\,000$) number of particles in Fig.\ \ref{fig:radialiceline}, in addition to $n_{\rm p}=10\,000$ particles. For particle sizes $R\lesssim0.1\,R_1$ we find converging results, whereas for larger particle sizes the growth is dominated by statistical fluctuations caused by runaway-growth of a few lucky particles. Our condensation model is constructed to give correct results in a regime where many particles compete for vapour, i.e.\ up to pebbles of a few decimeters, but may be unphysical in the regime of meter-sized boulders and beyond.

\subsubsection{Varying the radial extent of the simulation and the turbulent $\alpha$-value}
\label{sec:rad_2}

\begin{figure}
{\includegraphics[width=8.cm]{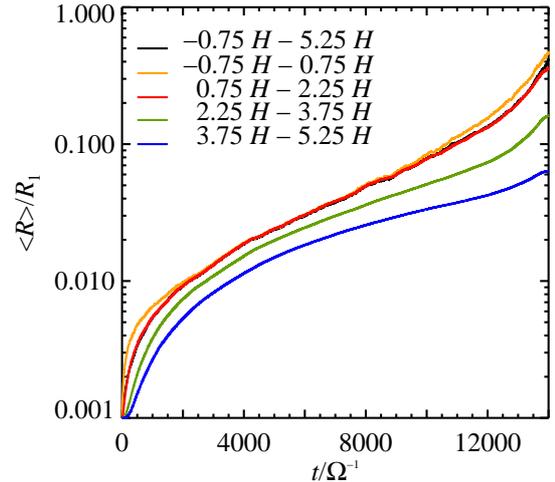}}
\caption{Growth in different radial subdomains of the simulation box shown in Fig.\ \ref{fig:rad_timeevo}. Black shows the average particle size in the total box, whereas yellow, red, green and blue gives the average particle sizes in the different subdomains, from left to right. The total particle size at any time is dominated by the growth near the radial ice line.}
\label{fig:partgrowth}
\end{figure}

\begin{figure*}
\includegraphics[width=18.0cm]{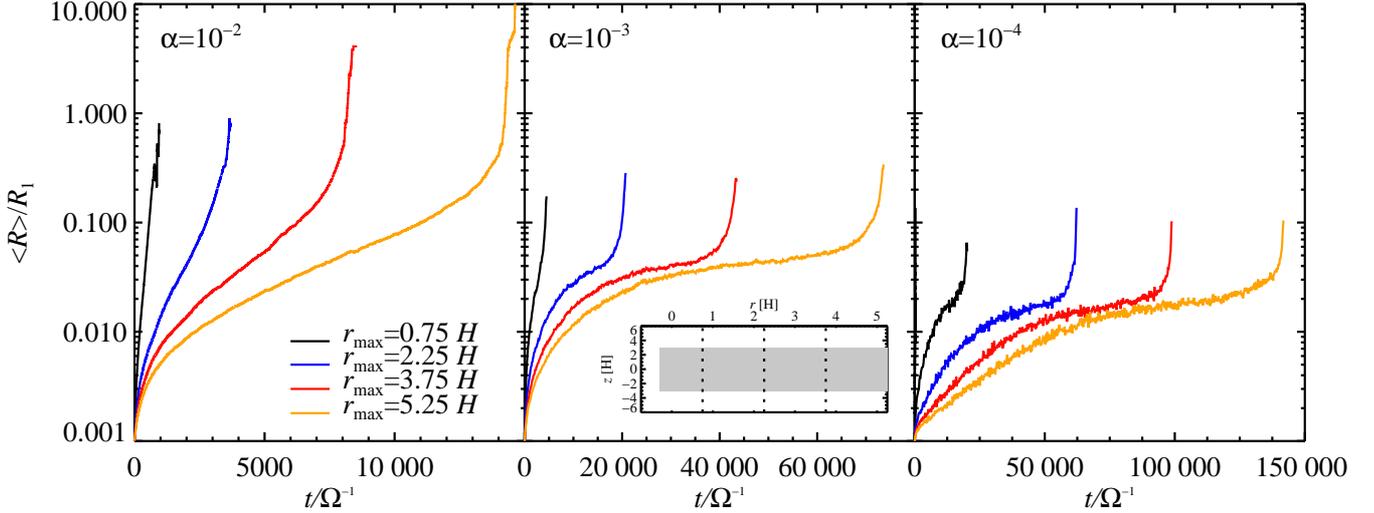}
\caption{Particle growth for different radial extents of the simulation box and  turbulence strengths. The mean particle size $\langle R \rangle$ is shown as a function of time in inverse Keplerian orbits $\varOmega^{-1}$ for $\alpha=10^{-2}$ (left panel), $\alpha=10^{-3}$ (middle panel) and $\alpha=10^{-4}$ (right panel). The size of the box has been extended in the radial direction, with $r_{\rm max}$ colour coded as: black 0.75 {\it H}, blue 2.25 {\it H}, red  3.75 {\it H} and yellow 5.25 {\it H}. For all runs $z_{\rm min}=-6.0\, H$,  $z_{\rm max}=6.0\, H$ and $r_{\rm min}=-0.75$ {\it H}. The inlay figure shows the simulation domain, with the considered radial extents marked.}
\label{fig:extbox_nopb}
\end{figure*}

To assess the importance of the size of the simulation domain, we ran simulations where the box size was varied in the radial direction. Fig.\ \ref{fig:rad_timeevo} shows the time evolution of a run where the radial extent of the simulation box has been increased, so that $r_{\rm max}=5.25\,H$. The most significant growth takes place close to the radial ice line, but due to radial mixing large particles can be found throughout the entire radial extent of the simulation box. In Fig.\ \ref{fig:partgrowth} the partial growth in four equally sized subdomains can be compared to the total growth in the whole simulation domain, as shown in Fig.\ \ref{fig:rad_timeevo}. This illustrates that the total particle growth is dominated by the growth near to the radial ice line.

The left panel of Fig.\ \ref{fig:extbox_nopb} shows the particle growth for different radial extents of simulation domains for a strongly turbulent disc, $\alpha=10^{-2}$. The growth timescale increases with larger simulation domains, as the amount of ice particles is increased while the supply of water vapour across the ice line remains unchanged. Results from simulations with weak turbulence are presented in the middle ($\alpha=10^{-3}$) and right ($\alpha=10^{-4}$) panels of Fig.\ \ref{fig:extbox_nopb}. These low-turbulence simulations show that, although slowly, particles can grow via condensation also in dead zones.

\section{Discussion}
\label{sec:discussion}

Our results show that ice condensation is an efficient way to form centimeter-sized and decimeter-sized pebbles. Ice condensation does not suffer from bouncing and fragmentation barriers and could be the dominant mode of growth in protoplanetary discs, to sizes where particles can concentrate strongly in the turbulent gas and continue towards planetesimal sizes by gravitational collapse. However, our results rely on a number of assumptions and simplifications which we discuss here.

\subsection{Other ice lines}

This model has been developed primarily to investigate the water ice line at $r\approx3\,\rm AU$. Similar ice lines, or condensation fronts, exist also for other chemical species, and as this model is scalable to any $r$ it can easily be adapted to include any condensation front. Of importance for planet formation purposes are in particular the ammonia, methane, carbon monoxide and molecular nitrogen ice lines further out in the protoplanetary disc, and the numerous silicate condensation fronts further in towards the central star. A global disc model, including all condensation fronts of the most important chemical species in the protoplanetary disc, will be an important extension of this work.

\subsection{Disc evolution}

The systematic motion of gas accretion onto the central star has been ignored in this work. However, compared to the time scales of particle growth by condensation the accretion process is slow enough to be safely neglected \citep{hartmannetal1998}. 

The ice line has throughout this work been considered as fixed. Seen over the life time of the disc this is not true, as the ice line probably moves both in a systematic way over long time scales and due to shorter heating events \citep{armitage2011}. However, as the time scale for growth by condensation found in this work is on the order of $\tau_{\rm c}\approx1000\,\varOmega^{-1}$, which is very short compared to the life time of the disc $\tau_{\rm disc}\sim3\,\rm Myr$, the assumption of a fixed ice line is reasonable. 

We also ignore the potential effect the processes of condensation and sublimation have on the ice line position, since the resulting release and absorption of energy is relatively small compared to the total thermal energy of the disc \citep{stevensonlunine1988}.

\subsection{Ice nuclei}
\label{sec:nuclei}

We have modelled ice particles as one-component particles; homogeneous water ice particles. In reality these particles would have (at least) a two-component composition, that of a rocky core and an ice mantle. This is due to the fact that supersaturated water vapour is needed for  homogeneous nucleation, i.e. for the spontaneous formation of a new ice particle. As this is typically not the case in the conditions prevailing in a protoplanetary disc, water vapour instead condenses onto an existing grain, either a rocky dust particle, or a particle with an already existing ice layer on top of the dust grain. Whether vapour preferably condenses onto a bare dust grain or an ice particle depends on material properties and on their respective sizes. 

Ignoring the small refractory dust particles present in the disc possibly leads to an over-estimation of the growth of the larger ice particles. For a fixed amount of mass, smaller dust particles have a larger combined surface area than larger particles, and therefore there is a higher probability that a vapour particle condenses onto a smaller dust superparticle than onto a larger superparticle. This is for the case when the material properties of dust and ice are such that two equally large dust and ice particles have equal probabilities of having a vapour particle condensing onto them. Instead of growth of a few large ice particles that sediment towards the midplane and thereby are protected against sublimation, there would thus be a continuous sublimation and condensation of the small ice particles at the ice line. This effect can already be seen in our condensation model, as small particles are more likely to cross the ice line and sublimate, and conversely, to grow by condensation, but could be amplified by the introduction of refractory grains.

Material effects might make this problem more severe. The dust grain composition is assumed to be similar to that of interstellar dust grains, either being pure interstellar dust grains or conglomerates thereof. This gives us the most important dust species as silicates and carbonaceous material, with silicates typically assumed to be the dominating one \citep{draine2003}.
Silicates have material properties that makes them more efficient as ice nuclei than pure water ice particles \citep{goumansetal2009}, implying that the effect of small particle growth at the expense of the larger ones is amplified when taking the material properties into account.

There are however a number of possible solutions to this problem, in terms of reasons for the vapour to condense onto ice particles instead of the small dust grains.
Firstly, the dust grains are not likely to all have the same size. Rather, they would have a size distribution where the smallest grains can be nanometer-sized \citep{draine2003}. The largest dust particles of relevance are the ones of the largest size that are strongly enough coupled to the gas to be present at the relevant vertical distance from the midplane. This is around micrometer-scale (see Fig.\ \ref{fig:condtest_nonrandom}). The Kelvin curvature effect states that the equilibrium vapour density for a curved surface is higher than for a flat surface, meaning that vapour more easily condenses onto a flat surface than a curved one, an effect which is most important for the smallest particles, around nanometer-sizes \citep{sirono2011}.  Therefore vapour will effectively prefer condensing onto the larger particles, leaving the smallest ones as bare dust grains.

Secondly, small grains are likely to be hotter than larger particles, moving the ice line of small particles closer to the midplane than the ice line of large particles. The temperature of the grains is set by the balance between absorption and emission efficiency. As grains absorb and emit radiation most efficiently in wavelengths smaller than and up to their own size, there is a size-dependent effect \citep{krivovetal2008, vitenseetal2012}. The spectral energy distribution of a solar-type star peaks at wavelengths of about $500\,{\rm nm}$. All grains of this size and larger therefore absorbs incoming radiation with approximately equal efficiency. However, grains emit in infra-red wavelengths (larger than 1 $\rm \mu m$), and thus the larger grains can emit more energy than the smaller ones. The resulting temperature for the smaller grains is therefore higher than for the larger grains. Also the material affects the temperature of the grains. As water ice is nearly transparent in visible light, where a solar-type star peaks, an ice particle is heated less by the stellar radiation than a particle of a more absorbing material \citep{lecaretal2006}. Both mechanisms are valid for grains residing in the atmosphere of the disc, where the disc is optically thin to the stellar radiation, but not for grains in the midplane, where the received stellar radiation is mainly the re-emitted infrared radiation from neighbouring grains.

Further, there is a possibility that material effects might prevent water vapour from condensing onto the dust grains. As mentioned above, bare silicate grains are more efficient as ice nuclei than ice-covered ones. However, for carbonaceous grains the opposite is true \citep{papoular2005}, meaning that vapour condenses more easily onto a water ice particle than onto a bare carbonaceous grain. For a grain population with ice coated particles and bare carbonaceous grains, the carbonaceous grain would therefore be left bare, whereas the ice grains would continue growing. 

\subsection{Particle collisions}
\label{sec:collisions}
We have neglected collisions between particles throughout this work. Depending on material properties and relative velocities of the particles, collisions can lead to increased growth via coagulation, to fragmentation or to bouncing. In general, small particles tend to stick together to a larger extent than large particles, if colliding, whereas larger particles are more prone to bouncing or fragmenting. Collisions amongst large particles therefore have a tendency to be destructive, whereas small-particle collisions are more likely to favour growth (or at least not counteract it). For silicate particles growth via collisions involving equal-sized particles is very difficult beyond millimeters, however collisional growth involving small particles colliding with a large target is possible, although slow \citep{wurmetal2005, johansenetal2008, windmarketal2012}.

Whereas extensive experimental data exists on collisions between rocky particles, 
the outcome of ice particle collisions is not equally well-known. \citet{bridgesetal1996} performed low-velocity collisions ($v_{\rm rel}<5\,\rm cm\,s^{-1}$) between ice pebbles and found that ice particles covered with a frost layer have an increased stickiness compared to rocky particles. A mechanism of collisional fusion, in which ice particles undergo a phase change during collision, has been suggested as a way for particles to stick also in collisions with higher velocities ($1 {\rm m\,s^{-1}} < v_{\rm rel}<100\,{\rm m\,s^{-1}}$) \citep{wettlaufer2010}.

Collision experiment for higher velocities have been used to derive criticial relative velocities above which fragmentation can occur, $v_{\rm crit}$. \citet{higaetal1996} found a critical velocity of $v_{\rm rel}\approx 1\,\rm m\,s^{-1}$, but other groups have found significantly larger values \citep{arakawa1999, arakawaetal2002}. Computer simulations suggest critical velocities of up to $v_{\rm rel}\approx 100\,\rm m\,s^{-1}$ \citep{benzasphaug1999}.

Collision velocities for millimeter-sized particles are set by the turbulent velocities and are expected to be on the order of a few $\rm m\,s^{-1}$, whereas collisions involving larger particles approach the radial drift velocity, $v_{\rm rel}\approx 50 \,{\rm m\, s^{-1}}$ \citep{braueretal2008}. Both coagulation and fragmentation are therefore expected to occur as a result of collisions. A likely scenario is that collisions between large particles lead to fragmentation, where the resulting small ice fragments are later swept up in collisions with other particles.

An important implication of collisions is that it provides a natural means of removing small dust grains released from the ice particles when sublimating. As these dust grains are very efficient ice nuclei, they might prevent growth of already large particles by ``stealing'' all the vapour, as discussed in Section \ref{sec:nuclei}. However, the small dust grains are likely to be swept up by larger ice particles in collisions. This does not increase the growth significantly, but has the important benefit that the small dust grains are removed so that vapour has to condense onto growing ice particles.

We plan to include the effects of multiple condensable species, refractory ice nuclei and particle collisions in future work. The present work highlights that simple turbulent dynamics can cause significant particle growth by condensation of volatiles, motivating follow-up studies with increasingly realistic models for the condensation and collision processes.

\section{Conclusions}
\label{sec:conclusions}

In this paper we have shown that ice condensation is an important particle growth mechanism that must to be taken into account in models of early planet formation. As the more thoroughly investigated mechanism of coagulation is inefficient in forming particles larger than centimeters, growth by condensation is an important mechanism that could complement coagulation or even be the dominant particle growth mode.

Our results show that growth by condensation near ice lines is rapid and results in large particle sizes. The growth time scale from millimeter-sized dust grains to decimeter-sized pebbles is $\tau_{c}\approx1000\,\rm years$. Significant growth is obtained for a range of turbulent $\alpha$-values from $10^{-4}$ to $10^{-2}$, where the higher value corresponds to the strength of turbulence that has been inferred from observations of protoplanetary discs, and the lower value has been suggested in a dead zone with weak turbulence stirred by the narrow active layers.

An implication of our model is a lower column density of vapour in the entire region outside the radial ice line compared to the inner part of the disc, caused by condensation onto existing grains and subsequent sedimentation of particles. This is in agreement with observations of CO \citep{qietal2011}, where a drop in abundance was found outside of the CO ice line. Similarly, the outer disc regions have been observationally inferred to be depleted in water vapour \citep{hogerheijdeetal2011}. Observed remnant water and CO vapour in the outer disc shows the importance of the coexistence of a cold midplane and an upper atmospheric layer where vapour can still exist.

Ice condensation can also explain observations of large quantities of pebbles in protoplanetary discs \citep{wilneretal2005}. Such pebbles are crucial in planet formation models, as they are the preferred starting size for planetesimal formation by clumping via streaming instabilities, in pressure bumps and in vortices, followed by gravitational collapse. Once large planetesimals have formed, continued pebble accretion is a very efficient path to formation of the cores of gas and ice giants \citep{ormelklahr2010, lambrechtsjohansen2012, morbidellinesvorny2012}. 

Ice condensation is a natural consequence oftemperature gradients in protoplanetary discs. With our simulations we have shown that condensation is an efficient growth mechanism with the potential to explain the formation of decimeter-sized pebbles in protoplanetary discs, thereby providing the missing link to further growth into planetesimals and planets.

\begin{acknowledgements}
We thank J{\"u}rgen Blum, Melvyn B. Davies, Ewine van Dishoeck, Kees Dullemond, Michiel Lambrechts, Klaus Pontoppidan, Erik Swietlicki, John Wettlaufer and Andrew Youdin for discussions and helpful suggestions. We also thank the referee, Phil Armitage, for comments that improved the quality of the paper. This research was partially funded by the European Research Council under ERC Starting Grant agreement 278675-PEBBLE2PLANET.
\end{acknowledgements}

\end{document}